\documentstyle[12pt]{article}

\begin{document}

\title{ \hfill {\large Preprint UGI-96-11\\
 \hfill Submitted to Z. Phys. C}\\[5mm]
Probing Hadronic Polarizations with Dilepton Anisotropies
\footnote{Supported by BMBF and GSI Darmstadt} \vspace*{5mm} \\}
\author{E.L. Bratkovskaya, W. Cassing and U. Mosel\vspace*{2mm} \\
\small \em Institut f\"ur Theoretische Physik, Universit\"at Giessen,
D-35392 Giessen, Germany}
\date{}
\maketitle

\begin{abstract}
We investigate the production of $e^+e^-$ pairs from $p + Be$ and
nucleus-nucleus collisions from 2 GeV/A to 200 GeV/A within a covariant
transport approach and focus on the dilepton angular anisotropies as a
function of the dilepton invariant mass. Whereas the low mass regime
yields information about the Dalitz decays of the $\Delta, \eta$ and
$\omega$, above M $\approx $0.5 GeV the angular anisotropies provide
additional information about the $\pi^+\pi^- \to \rho^0
\to e^+e^-$ channel in the medium. The anisotropy coefficient
is found to change its sign for M $>$ 0.5 GeV in case of
nucleus-nucleus reactions when increasing the bombarding energy from 2
GeV/A to 200 GeV/A which sheds some light on the $\pi \pi \to \rho$
dynamics.
\end{abstract}

\newpage
\section{Introduction}

Nowadays, dileptons are used as electromagnetic signals from the hot
and dense nuclear phase in heavy-ion collisions at BEVALAC/SIS
\cite{ro88,na89,ro89} or SPS energies~\cite{CERES200,CERES450,HELIOS}.
The information carried out by leptons may tell us not only about the
interaction dynamics of colliding nuclei, but also on properties of
hadrons in the nuclear environment or on a possible phase transition of
hadrons into a quark-gluon plasma (cf. \cite{Shyryak}).  However, there
are a lot of hadronic sources for dileptons because the electromagnetic
field couples to all charges and magnetic moments. In particular, in
hadron-hadron collisions, the $e^+e^-$ pairs are created due to the
electromagnetic decay of time-like virtual photons which can result from
the bremsstrahlung process or from the decay of baryonic and
mesonic resonances including the direct conversion of vector mesons
into virtual photons in accordance with the vector dominance
hypothesis.  In the nuclear medium, the properties of these sources may
be modified and it is thus very desirable to have experimental
observables which allow to disentangle the various channels of dilepton
production.

Apart from the differential $e^+e^-$ spectra the investigation of
lepton angular distributions is promising, too, because the virtual
photon created in hadronic interactions is polarized. The coupling of a
virtual photon to hadrons induces a dynamical spin alignment of both
the resonances and the virtual photons due to the conservation laws and
consequently the angular distribution of a lepton pair with respect to
the polarized photon momentum is anisotropic.

The angular distributions of leptons for large invariant masses (above
the $J/\Psi$ regime) were investigated at high energy, both
experimentally~\cite{ExperB&A} and theoretically (cf. the review of
Strojnowski~\cite{Strojnowski}), more than a decade ago. Here, the
dominant source at large invariant masses is the quark annihilation
(Drell-Yan) process.  The former measurements allowed to observe a
small deviation of the polar angular distribution from the prediction
of the naive quark-parton model, the deviations observed, however, are
in line with more detailed QCD computations.  Thus the sensitivity of
lepton angular characteristics already has been successfully used to
differentiate between models.

More recently, it has been proposed~\cite{BTT94,BSCMTT94} to use the
lepton pair angular distributions also for a distinction between
different sources in the "small" invariant mass region ($M<1$~GeV),
where a lot of dilepton sources contribute.  It has been shown,
furthermore,  that due to the spin alignment of the virtual photon and
the spins of the colliding or decaying hadrons, this {\em lepton decay
anisotropy} turns out to be sensitive to the specific hadronic
production channel. In our previous works
\cite{BTT94,BSCMTT94,BCMTTT95} we have calculated the dilepton
anisotropy for elementary nucleon-nucleon collisions and for $p + d$
reactions at BEVALAC energies. First results for proton-nucleus and
nucleus-nucleus collisions at 1~--~2~GeV/A are reported in \cite{BCM96},
where the calculated inclusive dilepton spectra were compared
to the DLS data~\cite{ro88,na89,ro89}.

In this work we present a more systematic study on the dilepton angular
anisotropy for $p + A$ and $A + A$ reactions from 2 GeV/A to 200 GeV/A
within the covariant transport approach HSD \cite{Eheh&Cas95}. At SPS
energies the inclusive dilepton spectra can be compared
to the results of the CERES/NA45 and HELIOS-3
collaborations~\cite{CERES200,CERES450,HELIOS}, a task that already has
been addressed with some success in \cite{Cassing95,Cass96}.  Here we
proceed with predictions for the corresponding dilepton angular
distributions and for the system $Au + Au$ at 10 GeV/A and 160 GeV/A.
Our work is organized as follows: in Section 2 we describe the
evaluation of the lepton pair angular distribution and briefly repeat
the main ingredients of the covariant transport approach. Section 3
contains the calculated results on dilepton spectra for $p + Be$ and
various nucleus-nucleus collisions as well as the dilepton anisotropies
as a function of the invariant dilepton mass. We summarize our study in
Section 4 and present a brief discussion on the experimental
requirements to measure the effects proposed.

\section{Theoretical ingredients}

\subsection{Definition of the anisotropy coefficient}

The general form of the angular distribution for the decay of a virtual
photon into a lepton pair may be written \cite{Got&Jack64} as
\begin{eqnarray}
W(\theta,\varphi)={3\over 8\pi}&&\hspace*{-6mm}
\left[ \rho_{11} (1+\cos^2\theta) + (1-2\rho_{11})\sin^2\theta
+ \rho_{1-1}\sin^2\theta \cos^2 2\varphi\right.
\nonumber\\
&&\left. +\sqrt{2} {\rm Re}\rho_{10}\sin 2\theta\cos\varphi\right],
\label{Wthetaphi}\end{eqnarray}
where the angles $\theta,\varphi$ of the electron momentum are measured
with respect to a fixed $z$-axis in the virtual photon rest frame. The
density matrix elements  $\rho_{ij}$ depend on the choice of the
reference frame as well as all the variables describing the virtual
photon.  An integration over the azimuthal angle gives
\begin{eqnarray}
W(\theta) \sim 1 + B \cos^2\theta,\label{Wtheta}
\end{eqnarray}
where the coefficient $B$ may vary between $-1$ and $+1$. Adopting the
normalization condition $\rho_{00} + 2\rho_{11} =1$, the coefficient
$B$ can be represented as (cf. \cite{Vasavada77}):
\begin{eqnarray}
B={3\rho_{11}-1\over 1-\rho_{11}}.
\label{B-rho}\end{eqnarray}

Adopting the general form Eq.~(\ref{Wthetaphi}) the angular distribution of
dileptons created in channel $i$ in a hadron-hadron ($h+h$)
reaction then can be written as~\cite{BTT94,BSCMTT94}
\begin{eqnarray}
{d\sigma_i^{hh}\over dM d\cos\theta_{hh}}
= A_i^{hh} (1 + B_i^{hh} \cos^2\theta_{hh}).
\label{in1}\end{eqnarray}
Here $\theta_{hh}$ is defined by $\theta_{hh}=(\widehat{\vec l_-^*, \vec
v_q^{hh}})$ with the electron momentum $\vec l_-^*$  measured in the
dilepton center-of-mass system ($\vec q^* \equiv \vec l_-^* + \vec
l_+^* = 0$), while $\vec v^{hh}_q= {\vec q}^{hh}/q_0^{hh}$ is the
velocity of the dilepton c.m.s. relative to the $h + h$ c.m. system.

The coefficient $B_i^{hh}$ in Eq.~(\ref{in1}) describes the anisotropy
of the angular distribution ($B_i^{hh} = 0$ in case of isotropy),
while $A_i^{hh}$ determines the magnitude of the respective cross section.
The absolute value of $B_i^{hh}$ now depends on the choice of the
coordinate system. For example, in the Gottfried-Jackson frame the
$z$-axis coincides with the direction of the incident beam in the
dilepton rest frame; this definition was adopted in the experiments
before~\cite{ExperB&A}.  Following~\cite{BTT94,BSCMTT94} we will use
for our analysis the "helicity" system, where the $z$-axis is chosen
along the direction of the virtual photon momentum $\vec q$ in the
center-of-mass system (c.m.s.) of the colliding hadrons.

The total differential cross section for $h+h$ collisions now
can be represented as a sum of the differential cross sections for
all channels $i$,
\begin{eqnarray}
{d\sigma^{hh} \over dM d \cos\theta_{hh}} = \sum\limits_{i=channel}
{d\sigma_i^{hh}\over dM d\cos\theta_{hh}} =
A^{hh}(M) (1+B^{hh}(M)\cos^2\theta_{hh}),
\label{Ssum}\end{eqnarray}
which leads to the total anisotropy coefficient,
\begin{eqnarray}
 B^{hh}(M) = \sum\limits_{i=channel} <B_i^{hh}(M)>,  \hspace*{1cm}
 <B_i^{hh}(M)> = {\displaystyle  {\displaystyle d\sigma_i^{hh}\over dM}
 \cdot {\displaystyle B_i^{hh}\over 1+{\displaystyle 1\over 3} B_i^{hh}}
\over \displaystyle \sum\limits_i {\displaystyle d\sigma_i^{hh}\over dM}
\cdot {\displaystyle 1\over 1+{\displaystyle 1\over 3} B_i^{hh}}},
\label{B_isum}\end{eqnarray}
where the special weighting factors originate from the necessary
angle-integrations.
Thus, the anisotropy coefficient $B^{hh}$ for $h+h$ reactions is the
sum of the ``weighted'' anisotropy coefficients ($<B_i^{hh}>$) for each
channel $i$ obtained by means of the convolution of $B_i^{hh}$ with the
corresponding invariant mass distribution (cf. \cite{BCMTTT95}).

The anisotropy coefficients for the bremsstrahlung, Dalitz decays of
$\Delta$-resonance and $\eta$-meson in $pn$ and $pp$ interactions have
been calculated on the basis of a one-boson-exchange model fitted to
elastic $NN$ scattering in Ref.~\cite{BSCMTT94}.  These anisotropy
coefficients are a function of the invariant mass $M$, the masses $m_a,
m_b$ and the initial invariant energy $\sqrt{s}$ of the hadrons $a + b$
involved in the reaction.  As shown in Ref.~\cite{BTT94}, the
anisotropy coefficient for pion annihilation in the $\pi^+ \pi^-$ c.m.s. is
given by $B_{\pi^+\pi^-} = -1$.  For this particular situation, where
there are only two particles in the initial and final states,
$\theta_{hh}$ is the angle of the lepton momentum with respect to the
pion momentum  in the c.m.s. of the leptons (or pions), i.e. ${\vec
 q}^{*}=\vec p_a^*+\vec p_b^* = \vec l_+^* +\vec l_-^* = 0$.

\subsection{Covariant transport approach for nucleus-nucleus collisions}

The dynamical evolution of proton-nucleus or nucleus-nucleus
collisions is described by a coupled set of transport equations
evolving phase-space distributions
$f_{h} (x,p)$ for each hadron $h$ \cite{Eheh&Cas95},
 i.e.\
\begin{eqnarray}  \label{g24}
\lefteqn{\left\{ \left( \Pi_{\mu}-\Pi_{\nu}\partial_{\mu}^p U_{h}^{\nu}
-M_{h}^*\partial^p_{\mu} U_{h}^{S} \right)\partial_x^{\mu}
+ \left( \Pi_{\nu} \partial^x_{\mu} U^{\nu}_{h}+
M^*_{h} \partial^x_{\mu}U^{S}_{h}\right) \partial^{\mu}_p
\right\} f_{h}(x,p) } \nonumber \\
&& = \sum_{h_2 h_3 h_4\ldots} \int d2 d3 d4 \ldots
 [G^{\dagger}G]_{12\rightarrow 34\ldots}
\delta^4_{\Gamma}(\Pi +\Pi_2-\Pi_3-\Pi_4 \ldots )  \nonumber\\
&& \times \left\{ f_{h_3}(x,p_3)f_{h_4}(x,p_4)\bar{f}_{h}(x,p)
\bar{f}_{h_2}(x,p_2)\right.  \nonumber\\
&& -\left. f_{h}(x,p)f_{h_2}(x,p_2)\bar{f}_{h_3}(x,p_3)
\bar{f}_{h_4}(x,p_4) \right\} \ldots\ \ .
\end{eqnarray}
In Eq.~(\ref{g24}) $U_{h}^{S}(x,p)$ and
$U_{h}^{\mu}(x,p)$ denote the real part of the scalar and vector
hadron selfenergies, respectively, while $[G^+G]_{12\rightarrow
34\ldots} \delta^4 (\Pi +\Pi_2-\Pi_3-\Pi_4\ldots )$ is the 'transition
rate' for the process $1+2\rightarrow 3+4+\ldots$ which is taken to be
on-shell in the semiclassical limit adopted.
 The hadron quasi-particle
properties in (\ref{g24}) are defined via the mass-shell
constraint \cite{Weber1},
\begin{equation}   \label{g25}
\delta (\Pi_{\mu}\Pi^{\mu}-M_{h}^{*2} ) \ \ ,
\end{equation}
with effective masses and momenta given by
\begin{eqnarray}
M_{h}^* (x,p)&=&M_h + U_h^{S}(x,p) \nonumber \\
\Pi^{\mu} (x,p)&=&p^{\mu}-U^{\mu}_h (x,p)\ \ ,
\label{g26}   \end{eqnarray}
while the phase-space factors
\begin{eqnarray}
\bar{f}_{h} (x,p)=1 \pm f_{h} (x,p)
\label{fxp}\end{eqnarray}
are responsible for fermion Pauli-blocking or Bose enhancement,
respectively, depending on the type of hadron in the final/initial
channel. The dots in Eq.~(\ref{g24}) stand for further contributions
to the collision term with more than two hadrons in the final/initial
channels. The transport approach (\ref{g24}) is fully specified by
$U_{h}^{S}(x,p)$ and $U_{h}^{\mu}(x,p)$ $(\mu =0,1,2,3)$, which
determine the mean-field propagation of the hadrons, and by the
transition rates $G^\dagger G\,\delta^4 (\ldots )$ in the collision
term, that describe the scattering and hadron production/absorption rates.

In the HSD\footnote{Hadron-String-Dynamics} approach the following
baryon states are explicitly propagated: the nucleon, $\Delta$,
$N^*(1440), N^*(1535), \Lambda, \Sigma$ and $\Xi$ as well as their
antiparticles with proper isospin. The meson sector includes all
pseudo-scalar mesons as well as the $1^-$ vector-meson octett. For the
baryon selfenergies we use the covariant parametrizations described in
detail in \cite{Eheh&Cas95}, whereas all mesons ($m$) in this work will
be propagated without selfenergies, i.e. $U^S_m = U^\mu_m = 0$. The
inelastic transition rates are as described in Ref.~\cite{Eheh&Cas95},
i.e. for invariant collision energies $\sqrt{s} < 2.6 $ GeV we employ
explicite cross sections for baryon-baryon reactions as in the BUU
approach of Ref.~\cite{Wolf} whereas for $\sqrt{s} \geq 2.6$ GeV we
employ the LUND string formation and fragmentation model \cite{lund}
to describe the hadron production cross sections. As in \cite{Eheh&Cas95}
a string formation time $\tau$ = 0.8 fm/c has been used to translate the
elementary 'free' cross sections to hadron reaction rates. In case of meson 
 -- baryon reactions explicite cross sections -- in line with the available
experimental data -- are taken into account for $\sqrt{s} < $ 1.8 GeV
whereas the LUND model is used above 1.8 GeV invariant collision energy.
Meson -- meson reactions, furthermore, are treated as binary collisions
including Breit-Wigner resonance cross sections with resonance properties
taken from \cite{PDT}.

The coupled set of transport equations (\ref{g24}) is solved within 
the testparticle method employing the parallel emsemble algorithm 
\cite{Cass90}, where a hadron is represented by a single testparticle 
in each individual parallel ensemble. In this way the 4-momenta of 
all hadrons are known throughout the collision together with their 
inelastic and elastic transitions. We note that the HSD approach 
quite well describes the reaction dynamics for $p + A$ and $A + A$ 
collisions  from SIS to SPS energies as demonstrated in 
\cite{Eheh&Cas95}.

The production of dilepton pairs $(e^+e^-)$ includes the Dalitz decays
$\Delta \to N e^+e^-, \eta \to \gamma e^+e^-, \omega \to \pi^0 e^+e^-$,
the bremsstrahlung for charged $NN, \pi N$ and $\pi \pi$ collisions as
well as the direct decays of the vector mesons $\rho, \omega$ and
$\phi$. Furthermore, the secondary mesonic channels $\pi^+\pi^- \to
\rho^0 \to e^+e^-, K \bar{K} \to \phi \to e^+e^-$ and $\pi \rho \to
\phi \to e^+e^-$ are taken into account, too.
The decay ratios are taken from the particle data
table \cite{PDT} whereas the formfactors for the Dalitz decays are
adopted from Landsberg \cite{Landsberg}.  Further details of the
computations are given in Refs.~\cite{Wolf,Cassing95,Cass96} and don't
have to be repeated here.

\subsection{Anisotropy coefficients for $p + A$ and $A + A$ reactions}

For heavy-ion reactions the situation becomes more complicated due to
the nuclear dynamics and the explicit time evolution of the interacting
system.  For the calculation of the nuclear anisotropy coefficient we
start from the point that the form of the angular distribution for all
"elementary" hadron-hadron interactions $a+b$, that occur in the
nucleus-nucleus reaction $A+B$, are known from the microscopic
calculations~\cite{BTT94,BSCMTT94,BCMTTT95}.

In case of A + A reactions we have to take into account that
the "elementary" $a+b$ c.m.s. (in
which the $B_i^{ab}$ are computed) are moving relative to the $A+B$
c.m.s.  This implies that the direction of the virtual photon momentum
$\vec q^{ab}$ relative to the hadron-hadron c.m.s. $a+b$ is different
for each elementary interaction $a+b$ and can't be restored from the
experimental data.  However, the direction of the virtual photon
momentum $\vec q^{AB}$ in the nucleus-nucleus c.m.s. $A+B$ can be
reconstructed for each lepton pair.  Thus, it is necessary to perform
an angular transformation from $\theta_{ab}$ to $\theta_{AB}$, where
$\theta_{AB}$ is the angle between the lepton momentum $\vec l^*$ and
the velocity of the dilepton c.m.s.  relative to the c.m.s. of the
colliding nuclei $A+B$ ($\vec v_q=\vec q^{AB}/q_0^{AB}$).

At high energies ($\sim$ 200 GeV/A) the main contributions to the 
dilepton spectra come from the Dalitz decays of $\eta, 
\omega$-mesons, $\pi^+\pi^-$-annihilation and direct decay of vector 
mesons $\rho,\omega,\phi$. As shown in Ref.~\cite{Cassing95} the 
contributions from the $\Delta$-Dalitz decay and bremsstrahlung are 
negligible.  Furthermore, the polarization of vector mesons created 
in direct nucleon-nucleon reactions is practically zero according to 
the measurements of Blobel et al.~\cite{Blobel}. This is due to the 
fact that the exclusive cross section $NN \to NN \rho^0,\omega,\phi$ 
is very small compared to the inclusive cross section $NN \to NN 
\rho^0, \omega, \phi + X$ containing several pions in the final 
channel, too.  Consequently, we do not consider the contribution to 
the anisotropy coefficient from the decays of vector mesons produced 
in primary $NN$-collisions.  Thus, there are only three dominant 
channels at intermediate and high energies, i.e. the Dalitz decays of 
$\eta$ - and $\omega$ - mesons as well as $\pi^+\pi^-$ annihilation 
that  have to be taken into account in our analysis.

For the computation of the nuclear anisotropy coefficient we use
another algorithm than in our previous paper~\cite{BCM96}, i.e. we
simulate explicitly lepton events for each channel $i$ with fixed
angular distribution in the individual $a+b$ system and then transform
the lepton and photon 4-momenta to the $A+B$ system.  For example, for
the Dalitz decay of the $\eta$-meson, the lepton angular distribution
in the rest frame of the $\eta$-meson has a form (\ref{Wtheta}) with
$B_\eta^{ab}=+1$ \cite{BTT94}, where $\theta_{ab}$ is the angle between
the lepton momentum $\vec l^*$ and the direction of $\vec q^\eta$,
while $\vec q^\eta$ is the virtual photon momentum in the rest frame of
the $\eta$.  We also use that the distribution of the virtual photon
momentum $\vec q^\eta$ in the rest frame of the $\eta$ is isotropic.
Thus, for each $\eta$ event with energy $E_\eta$ and momentum $\vec P_\eta$
in the nucleus c.m.s. $A+B$ we generate lepton events distributed with
respect to $\vec q^\eta$ according to Eq.~(\ref{Wtheta}) with
$B_\eta^{ab}=+1$ in the lepton c.m.s.
The next step then is the Lorentz transformation from the $\eta$
rest frame to the c.m.s. of $A+B$:
$\vec q_{AB} = L\left({\vec P_\eta \over E_\eta}\right) \vec q_\eta$,
where ${q_0}_\eta = {m_\eta^2+M^2\over 2 m_\eta}, \ |\vec q_\eta| =
\sqrt{\lambda(m_\eta^2,M^2,0)\over 2 m_\eta}$.
Finally we compute the angle $\theta_{AB}$ between  $\vec l^*$ and
$\vec v_q^{AB}$ as
\begin{eqnarray}
\cos\theta_{AB} ={\vec l^* \cdot \vec v_q^{AB}\over
|\vec l^*| \cdot |\vec v_q^{AB}|},
\label{cosAB}\end{eqnarray}
where $\vec v_q^{AB} =\vec q^{AB}/q_0^{AB}$ and $|\vec l^*|=M/2$ for
fixed invariant mass $M$.
Thus the angular distribution as a function of $\cos\theta_{AB}$ for all
generated lepton events is recovered and  $B_\eta$ can be computed via
\begin{eqnarray}
B_\eta^{AB} ={W(\theta_{AB}=0^o)\over W(\theta_{AB}=90^o)} -1.
\label{Beta}\end{eqnarray}
The weighted $<B_\eta>$ then follows from Eq.~(\ref{B_isum}).

In a similar way we calculate the anisotropy coefficient for the pion 
annihilation channel by using that the anisotropy coefficient in the 
c.m.s. of the leptons (or pions)~\cite{BTT94} 
$B_{\pi^+\pi^-}^{ab}=-1$.  For each  pion annihilation event we then 
restore the virtual photon energy and momentum in the c.m.s. of 
$A+B$:  $q_0^{AB} = E_a+E_b, \ \vec q^{AB} = \vec p_a + \vec p_b$, 
where $E_a, E_b$, $\vec p_a, \vec p_b$ are the energies and momenta 
of the pions in the $A+B$ c.m.s.  Then we perform a Lorentz 
transformation for the pion momentum from the $A+B$ c.m.s.  to the 
leptons c.m.s.:
$\vec p_a^* = L\left({\vec q^{AB} \over q_0^{AB}}\right) \vec p_a$.
In the lepton c.m.s. we generate the lepton events with
$|\vec l^*|=M/2$ ($ M^2={q_0^{AB}}^2- \vec q^{{AB}^2} $)
and angular distribution $W(\theta_{ab}) = 1 -\cos^2\theta_{ab}$,
where $\theta_{ab}$ is the angle between the $\vec l^*$ and
the pion momentum $\vec p_a^*$ in the leptons c.m.s.;
$\cos\theta_{ab} = {\vec l^* \cdot \vec p_a^* \over |\vec l^*|\cdot |\vec
p_a^*|}$. For each selected lepton event the angle $\theta_{AB}$ can be
calculated according to Eq.~(\ref{cosAB}).  The weighted
coefficient $<B_{\pi^+\pi^-}>$ can be computed in the similar manner as
in the $\eta$ case.

The anisotropy coefficient for the Dalitz decay of the $\omega$-meson,
furthermore, is computed in analogy to the $\eta$ case, using the
elementary coefficient $B_\omega^{ab} =+1$.

We have tested our algorithm for BEVALAC energies and -- within the
statistical accuracy --
reproduce the results from our previous paper~\cite{BCM96}, where an
approximation was made to calculate the anisotropy coefficients for the
Dalitz decays of $\Delta$ and $\eta$:  in~\cite{BCM96} we have used
the results of our microscopic calculations for the anisotropy
coefficient that were already averaged  over $\vec q^{ab}$ whereas, in
principle,  we need this degree of freedom to transform correctly the
differential cross section from the $a+b$ to the $A+B$ c.m.s.  Our
discrete algorithm now allows to avoid this approximation; it is close
to an experimental event by event analysis and includes the full hadron
dynamics.

\section{Numerical results for $p + Be$ and $A + A$ reactions}

In this section we present the results of our calculations for the
dilepton spectra and the anisotropy coefficient for the systems $p+Be$,
$Ca+Ca$, $S+Au$, $Au+Au$ from BEVALAC to SPS energies without employing
any medium effects for the mesons, i.e. $U_m^S=0, \ U_m^\mu=0$ in the
set of transport equations (\ref{g24}) for all mesons.

In the Fig.~\ref{fig1} we show the computed weighted anisotropy
coefficients $<B_i(M)>$ for $p+Be$ and $Ca+Ca$ collisions at the
bombarding energy of 2~GeV/A essentially confirming our previous
results from~\cite{BCM96} employing the averaged coefficients as
described above. We note that the inclusive dilepton spectra for the
reactions have been presented in Ref.~\cite{BCM96} together with
a "cocktail" decomposition. The main contributions arise from the $\eta$
and $\Delta$ Dalitz decays due to their large 'elementary' anisotropy
coefficients and cross sections, respectively. The contribution from
$pn$ bremsstrahlung is practically zero at all invariant masses due to a
small 'elementary' anisotropy coefficient.
The weighted coefficient from $\pi^+\pi^-$
annihilation is rather small ($\approx$ 0.1) even for the $Ca  + Ca$
reaction and decreases for $M \geq m_\rho$ due to the threshold
behaviour of the cross section. However, compared to
$<B_{\pi^+\pi^-}(M)> $ for $p+Be$, where pion annihilation is very low,
a clear (but moderate) enhancement can be extracted.

Before going over to the presentation of the anisotropy coefficient at
AGS energies  ($\sim 10$~GeV/A) we show the differential multiplicity
$dn_{e^+e^-}/dM$ for $p+Be$ and $Au+Au$ in order to demonstrate the
relative contribution of the different channels employing an 'energy
resolution' of 50 MeV (Fig.~\ref{fig2}).  In
the 'cocktail' plot the $\omega\to\pi e^+e^-$ Dalitz decay, $\eta$
Dalitz decay ('$\eta$'), the pion annihilation channel ('$\pi^+\pi^-$') as
well as the direct decay of the vector mesons ('$\rho,\omega,\Phi$')
are displayed explicitly while the solid curves (denoted by '$all$')
represent the sum of all sources.  The contributions from the $\Delta$
Dalitz decay and the bremsstrahlung channels are low by an order of magnitude
compared to the $\eta$-channel
and not plotted explicitly.  Similar to the systems at BEVALAC energies
the $\pi^+\pi^-$ annihilation channel is of no importance for $p + Be$,
but becomes the most significant sources for $M > 0.4$~GeV in case of
$Au + Au$ collisions.  Furthermore, when scaling the spectra for
$Au+Au$ by the number of projectile nucleons, we observe an additional
 enhancement in the $\Phi$-mass region which is due to the meson
induced secondary collisions channels $K \bar{K} \to \Phi \to e^+e^-$
and $\pi\rho \to \Phi \to e^+e^-$.

In Fig.~\ref{fig3} we show the weighted anisotropy coefficients
$<B_i(M)>$ for $p+Be$ and $Au+Au$ collisions at 10 GeV/A using the same
notations as in Fig.~\ref{fig2}.  The main contribution arises from the
$\eta$-Dalitz decays whereas the contribution from the $\omega$-Dalitz
decay is quite small due to its reduced cross section. The weighted
anisotropy coefficient from pion annihilation for $p+Be$ is zero for
the same reason. However, the $<B_{\pi^+\pi^-}(M)>$ for $Au+Au$ is zero
because the angular distribution of the annihilating pions relative to
the $\vec q^{AB}$ direction becomes practically isotropic at this
energy which leads automatically to an isotropic dilepton spectrum.
Obviously, this effect is related to our choice of the $z$-axis along
the $\vec q^{AB}$. One can show, for example, that in the
Gottfried-Jackson frame, the distribution of annihilating pions is
anisotropic and $<B_{\pi^+\pi^-}(M)>$ is small and negative at this
energy.

Before presenting our results for SPS energies we like to point out
that the transport approach HSD provides a good reproduction of the
pion rapidity and transverse energy distribution at SIS/BEVALAC and AGS
energies \cite{Eheh&Cas95} which ensures that the actual pion
phase-space densities and thus the $\pi^+\pi^-$ annihilation rate
should be realistic. However, a definite proof for SPS is still
lacking, since only rapidity distributions have been presented so far
in comparison to the experimental data for $S + Au$ at 200 GeV/A in
\cite{Cassing95}. We thus show in Fig.~\ref{fig4} the calculated
invariant cross section for $\pi^0$ mesons in the rapidity interval
$2.1 \le y \le 2.9$ as a function of the transverse mass of the pions
for $S+Au$ at 200 GeV/A in comparison  with the experimental data of
the WA80 collaboration~\cite{WA80_95}.  As can be seen from
Fig.~\ref{fig4} the results of our computation (open circles) are in a
good agreement with the experimental data (full circles) also with respect
to the transverse mass distribution.

The dilepton invariant mass spectra for $p+Be$ at 450 GeV/A and $S+Au$
at 200 GeV/A have been presented in \cite{Cassing95}.  For
'free' meson masses and formfactors the experimental cross section is
slightly underestimated for invariant masses $0.3 \le M \le 0.45$~GeV,
which may be due to a dropping of the $\rho$-meson mass at finite
baryon density \cite{Cassing95,li95}. In the present work we do not
explore this idea in more detail and calculate the dilepton spectra
without any medium modifications.

In Fig.~\ref{fig5} we show the differential dilepton multiplicities for
$Au+Au$ at 160 GeV/A, a system that has recently been investigated
experimentally at the SPS.  Again the dominant contributions come from
the $\eta$ Dalitz decay at small invariant masses and from the pion
annihilation and direct decay of vector mesons at $M\sim m_\rho$.

In Fig.~\ref{fig6} we present the weighted anisotropy coefficients
$<B_i(M)>$ for $p+Be$ collisions at 450 GeV/A, $S+Au$ at 200 GeV/A and
$Au+Au$ at 160 GeV/A. For $p+Be$ the situation is similar to the
previous cases at lower energies, the main contribution at small
invariant masses comes from the $\eta$ Dalitz decay;
$<B_{\pi^+\pi^-}(M)>$ is approximately zero due to the small cross
section. For nucleus-nucleus collisions, however, the contribution of
the pion annihilation channel becomes more essential.  As seen from
Fig.~\ref{fig6},  contrary to the lower bombarding energies, the weighted
anisotropy coefficient for pion annihilation is negative; an effect
related to some differences in the pion dynamics at low and high
energy as we will discuss below.

In order to explore this effect we use the result from \cite{BCM96}
that the dilepton anisotropy coefficient $<B_{\pi^+\pi^-}(M)>$ is
directly connected with the angular distribution of annihilating pions
$W(M,\cos\theta_\pi)$ relative to  $\vec q^{AB}$:
\begin{eqnarray}
B_{\pi^+\pi^-}^{AB} (M) ={ \int d\cos\theta_\pi \ \tilde A(\theta_\pi) \
\tilde B(\theta_\pi) \ W(M,\cos\theta_\pi)  \over
\int d\cos\theta_\pi \ \tilde A(\theta_\pi) \ W(M,\cos\theta_\pi)}.
\label{Bpi}\end{eqnarray}
The quantities $\tilde A(\theta_\pi), \tilde B(\theta_\pi)$
result from a Lorentz transformation from the angle $\theta_{ab}$
to $\theta_{AB}$ and read
\begin{eqnarray}
&& \tilde A(\theta_\pi) = 1 + {B_{\pi^+\pi^-}^{ab}\over 2} \sin^2\theta_\pi,
\hspace*{5mm}  \tilde B(\theta_\pi) = {B_{\pi^+\pi^-}^{ab}\over 2
\tilde A(\theta_\pi)} (3 \cos^2\theta_\pi - 1).  \label{tildeAB}
\end{eqnarray}
Here $B_{\pi^+\pi^-}^{ab}=-1$, while $\theta_\pi$ is the angle between
the pion momentum $\vec p_a^*$ in the c.m.s. of $a+b$ and the direction
of the vector $\vec v_q^{AB}$. In the $A+B$ c.m.s. $\theta_\pi$ can be 
written as
\begin{eqnarray}
\cos\theta_\pi = {M \ \Delta E \over  \sqrt{(M^2-4m_\pi^2) \ (E^2-M^2)}},
\label{cospi}\end{eqnarray}
where $\Delta E = E_a - E_b$, $E=E_a+E_b$
are the relative and total energies of the annihilating pions with
energy $E_a, E_b$ in the c.m.s. of $A+B$.

In Eq.~(\ref{cospi}) the $\pi^+\pi^-$ angular distribution
$W(M,\cos\theta_\pi)$ describes the energy and angular distribution
of the pions that annihilate.  Because of simple kinematical relations
the distribution of the annihilating pions  in $\Delta E$ and $E$
becomes broader with increasing bombarding energy. In line with
Eq.~(\ref{cospi}) this leads to a transition from a distribution
$W(M,\cos\theta_\pi)$ peaked at $90^o$ at low energies to a much
flatter one at high energies or to a positive $B$
coefficient at small energies and a negative one at high energies.

To quantify our above arguments for the coefficient
$<B_{\pi^+\pi^-}(M)>$ we show in  Fig.~\ref{fig7} the distribution
$W(M,cos\theta_\pi)$, a contour plot of $\cos\theta_\pi$ in the plane
$(\Delta E, E)$ at $M=0.7$~GeV (see Eq.~(\ref{cospi})), as well as
contour plots of the density distribution $W(\Delta E, E)$ for $Ca+Ca$
at 2 GeV/A, $Au+Au$ at 10 GeV/A and $S+Au$ at 200 GeV/A.  The value of
$\cos\theta_\pi$ is zero at $\Delta E=0$ and goes to $\pm 1$ when
increasing the absolute value of $\Delta E$ according to
Eq.~(\ref{cospi}). First of all one can see a pronounced change in the
shape of $W(M,\cos\theta_\pi)$ from $Ca + Ca$ at 2 GeV/A to isotropy for
$Au + Au$ at 10 GeV/A and to a forward -- backward peaked distribution for
$S + Au$ at 200 GeV/A. In the ($E,\Delta E$)-plane the maximum of
$W(\Delta E, E)$ is located near $\Delta E=0$ at low energy which
corresponds to $\cos\theta_\pi = 0$.  Thus, at low energy the
probability for pions to annihilate at angle $\theta_\pi=90^o$ relative
to the $\vec q^{AB}$ is larger than at $0^o,180^o$; correspondingly
$W(M,\cos\theta_\pi)$ has a maximum at $90^o$ (r.h.s). Furthermore, as
seen from Fig.~\ref{fig7} the absolute value of the $W_{max}(\Delta E, E)$
at $\Delta E = 0$ decreases with increasing energy and the
probability for annihilating pions with $\theta_\pi=0^o,180^o$
increases. In other words, at high energies  the annihilating pions
have larger longitudinal than transverse momentum in the c.m.s. of
leptons relative to $\vec q^{AB}$ (r.h.s.), and according to
Eq.~(\ref{Bpi}) the anisotropy coefficient $B_{\pi^+\pi^-}^{AB}$
becomes negative.

\section{Summary}

In this work we have presented a fully dynamical study of dilepton
production in proton-nucleus and nucleus-nucleus collisions from
SIS/BEVALAC to SPS energies. Due to the parallel ensemble algorithm
employed we have, furthermore, calculated dilepton angular anisotropies
on an event by event basis for the channels $\eta \to \gamma e^+e^-$,
$\omega \to \pi^0 e^+e^-$, $\Delta \to Ne^+e^-$ as well as dilepton
bremsstrahlung and $\pi^+\pi^-$ annihilation.  In all reactions we find
the anisotropy coefficient $<B>$ to be dominated by the $\eta$-decay
for invariant masses $M <$ 0.4 GeV, while $<B>$ is practically zero for
$M > 0.5$ in $p + Be$ reactions at all bombarding energies.  This situation
changes significantly  for nucleus-nucleus collisions showing a
positive $<B>$ from $\pi^+\pi^-$ annihilation of about 0.1 at 2 GeV/A,
turning approximately zero at 10 GeV/A and becoming negative $<B>
\approx -0.1$ at SPS energies. We could demonstrate that this energy
variation is due to the different pion annihilation dynamics.

Thus, the calculated anisotropy coefficients for nucleus-nucleus
collisions support our suggestion in
Refs.~\cite{BTT94,BSCMTT94,BCMTTT95} that the dilepton decay anisotropy
may serve as an additional observable to decompose the dilepton spectra
into the various sources and to obtain additional information on the
reaction dynamics. We note, however, that our transport calculations
correspond to the full acceptance for the lepton pairs. Especially cuts
in the transverse momenta affect the anisotropies sensitively. This
should be born in mind when analyzing experimental data with a limited
acceptance.

\section*{Acknowledgment}

The authors like to thank O.V. Teryaev and V.D. Toneev for the
cooperation on earlier stages of this work, their critical remarks and
valuable comments and discussions. Our special gratitude is to S.S.
Shimanskij for stimulating suggestions and encouraging talks. We are
grateful also to A.M. Baldin, A. Drees, A.A. Sibirtsev, S. Teis, A.I.
Titov and Gy. Wolf for clarifying discussions.

\newpage
\section*{Figure captions}

\begin{figure}[h]
\caption{
The weighted anisotropy coefficients $<B_i(M)>$ for $p+Be$ collisions
at the bombarding energy of 2.1  GeV and $Ca+Ca$ collisions
at the bombarding energy of 2.0 GeV/A.
The ``$\eta$'' denotes the contribution of the $\eta$--channel, the
``$\Delta$'' labels the contribution of the $\Delta$ Dalitz decay,
``$pn$''  the proton--neutron bremsstrahlung, and
``$\pi^+\pi^-$'' the pion annihilation channel.
The solid curves (denoted by ``$all$'') show the sum of all sources.}
\label{fig1}

\caption{
The differential multiplicity $dn_{e^+e^-}/dM$ for $p+Be$ and central 
$Au+Au$ collisions at a bombarding energy of 10 GeV/A.  The 
``$\omega\to\pi e^+e^-$" is the $\omega\to\pi e^+e^-$ Dalitz decay, 
``$\rho$'', ``$\omega$'', ``$\Phi$'' are the direct decays of the 
vector mesons.  The other notations are the same as in 
 Fig.~\protect\ref{fig1}. }
\label{fig2}

\caption{
The weighted anisotropy coefficients $<B_i(M)>$
for $p+Be$ and central $Au+Au$ collisions at 10 GeV/A.  The notation is the
same as in Fig.~\protect\ref{fig2}.  }
\label{fig3}

\caption{
The calculated invariant cross section (open circles) for $\pi^0$ mesons
in the rapidity interval $2.1 \le y \le 2.9$ as a function of the
transverse pion mass for $S+Au$ at 200 GeV/A in comparison  with the
experimental data (full circles) of the WA80
collaboration~\protect\cite{WA80_95}.  }
\label{fig4}

\caption{
The differential multiplicity $dn_{e^+e^-}/dM$ for a central $Au+Au$
collision at 160
GeV/A.  The notation is the same as in Fig.~\protect\ref{fig2}. }
\label{fig5}

\caption{
The weighted anisotropy coefficients $<B_i(M)>$ for $p+Be$ collisions
at 450 GeV/A, $S+Au$ at 200 GeV/A and $Au+Au$ at 160 GeV/A.
The notation is the same as in Fig.~\protect\ref{fig2}. }
\label{fig6}
\end{figure}

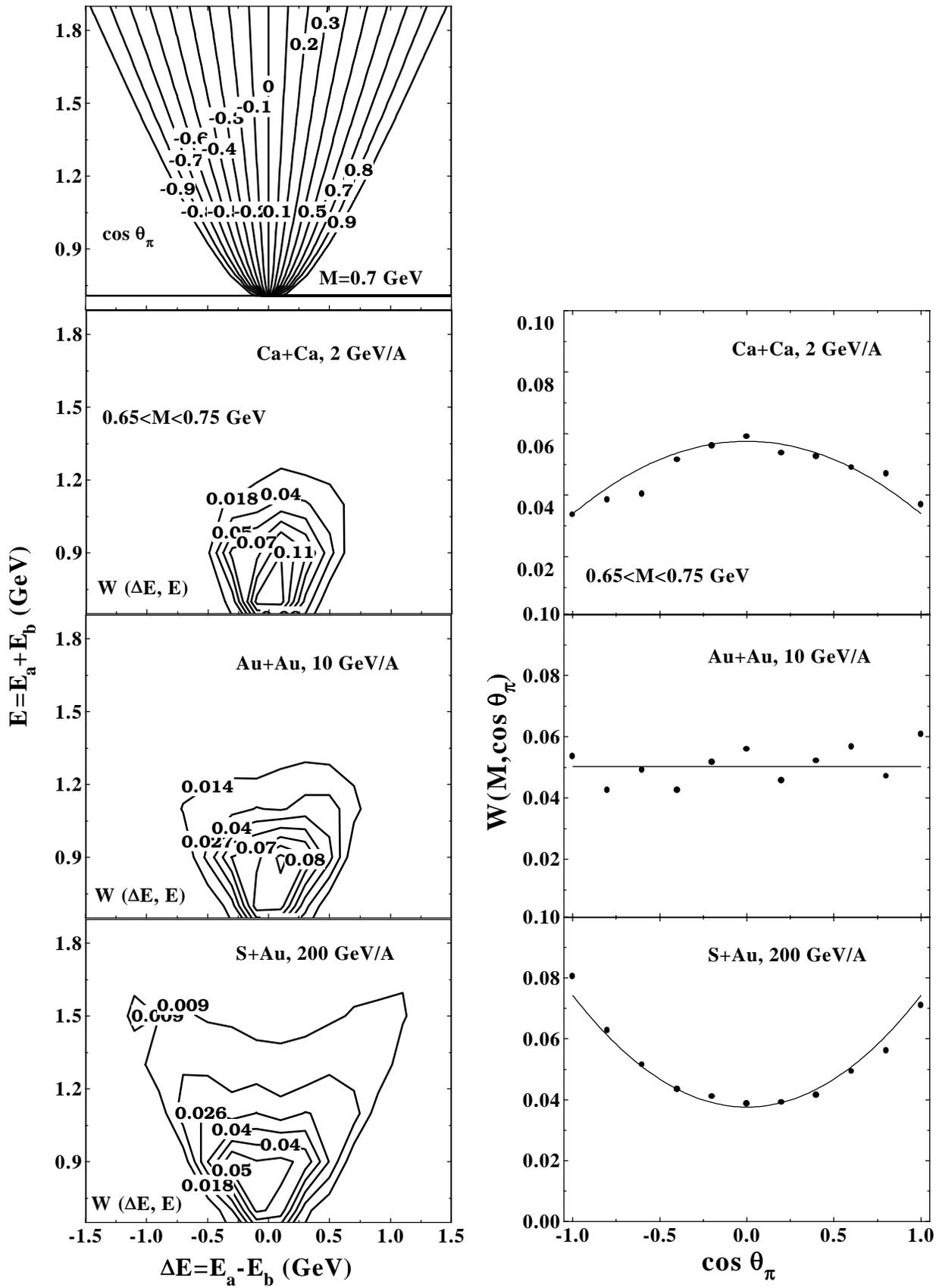
\begin{figure}[h]
\caption{
(l.h.s.): The contour plot of $\cos\theta_\pi$ in the plane $(\Delta 
 E, E)$ at $M=0.7$~GeV (see Eq.~(\protect\ref{cospi})), as well as 
the countor plots of the density distribution $W(\Delta E, E)$ for 
$Ca+Ca$ at 2 GeV/A, $Au+Au$ at 10 GeV/A and $S+Au$ at 200 GeV/A. 
(r.h.s.) The angular distribution $W(M,cos\theta)$ for $0.65 \le M 
\le 0.75 $ GeV for the same systems. The solid lines represent fits 
to the calculated distributions.}
\label{fig7}
\end{figure}


\newpage

\begin{thebibliography}{99}
\bibitem{ro88}
	G. Roche et al., Phys. Rev. Lett. {\bf 61} (1988) 1069
\bibitem{na89}
	C. Naudet et al., Phys. Rev. Lett. {\bf 62} (1989) 2652
\bibitem{ro89}
	G. Roche et al., Phys. Lett. {\bf 226B} (1989) 228
\bibitem{CERES200}
	G. Agakichiev et al., Phys. Rev. Lett. {\bf 75} (1995) 1272.
\bibitem{CERES450}
	T. Akesson et al., Z. Phys. {\bf C68} (1995) 47.
\bibitem{HELIOS}
	M.A. Mazzoni et al., Nucl. Phys. {\bf A566} (1994) 95c.
\bibitem{Shyryak}
	E.V. Shuryak, Phys. Rep. {\bf 61} (1980) 71.
\bibitem{ExperB&A}
	J. Badier et. al., Phys. Lett. {\bf B89} (1979) 145;
	K.J. Anderson et al., Phys. Rev. Lett. {\bf 42} (1979) 944.
\bibitem{Strojnowski}
	R. Strojnowski, Phys. Rep. {\bf 71} (1981) 1.
\bibitem{BTT94}
	E.L.~Bratkovskaya,  O.V.~Teryaev and V.D.~Toneev,
       Phys. Lett. {\bf B348} (1995) 283.
\bibitem{BSCMTT94}
	E.L.~Bratkovskaya, M.~Sch\"afer, W.~Cassing, U.~Mosel,
       O.V.~Teryaev, and V.D.~Toneev,
       Phys. Lett. {\bf  B348} (1995) 325.
\bibitem{BCMTTT95}
	E.~L.~Bratkovskaya,  W.~Cassing, U.~Mosel,
	O.~V.~Teryaev, A.~I.~Titov, and V.~D.~Toneev,
	Phys. Lett. {\bf  B362} (1995) 17.
\bibitem{BCM96}
	E.~L.~Bratkovskaya,  W.~Cassing, U.~Mosel, nucl-th/9601018,
	Phys. Lett. { \bf B} (1996), in press.
\bibitem{Eheh&Cas95}
	W. Ehehalt and W. Cassing, hep-ph/9507276,
	Nucl. Phys. { \bf A} (1996), in press.
\bibitem{Cassing95}
	W. Cassing, W. Ehehalt and C.M. Ko,
	Phys. Lett. {\bf  B363} (1995) 35.
\bibitem{Cass96}
	W. Cassing, W. Ehehalt and I. Kralik,
	Phys. Lett. {\bf  B} (1996) in press.
\bibitem{Got&Jack64}
	K. Gottfried and J.D. Jackson, Nuovo Cimento {\bf 33} (1964) 309.
\bibitem{Vasavada77}
	K.V. Vasavada, Phys. Rev. {\bf D16} (1977) 146.
\bibitem{Weber1}
	K. Weber et al., Nucl. Phys. {\bf A552} (1992) 713.
\bibitem{Wolf}
	Gy. Wolf, G. Batko, W. Cassing, U. Mosel, K. Niita,
	and M. Sch\"afer, Nucl. Phys. {\bf A517} (1990) 615;
	Gy. Wolf, W. Cassing and U. Mosel,
	Nucl. Phys. {\bf A552} (1993) 549;
\bibitem{lund}
	B. Nilsson-Almqvist and E. Stenlund,
	Comp. Phys. Comm. {\bf 43} (1987) 387.
\bibitem{PDT}
	Review of Particle Properties, Phys. Rev. {\bf D50} (1994) 1173.
\bibitem{Cass90}
	W. Cassing, V. Metag, U. Mosel, and K. Niita,
	Phys. Rep. {\bf 188} (1990) 363.
\bibitem{Landsberg}
	L.G. Landsberg, Phys. Rep. {\bf 128} (1985) 301.
\bibitem{Blobel}
	V. Blobel et al., Phys. Lett. {\bf B48} (1974) 73.
\bibitem{WA80_95}
	R. Albrecht et al., Phys. Lett. {\bf B361} (1995) 14.
\bibitem{li95}
	G.Q. Li, C.M. Ko and G.E. Brown,
	Phys. Rev. Lett. {\bf 75} (1995) 4007.
\end{thebibliography}
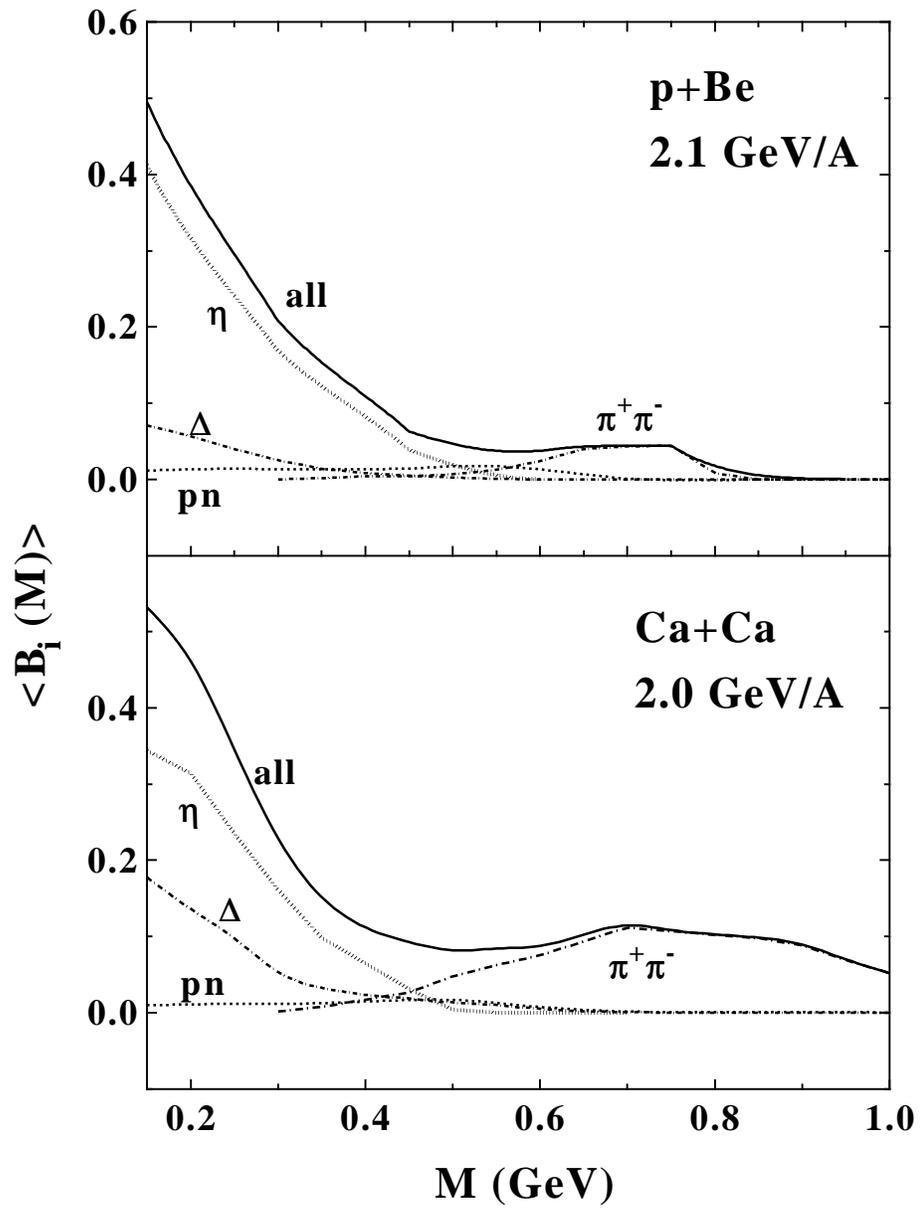
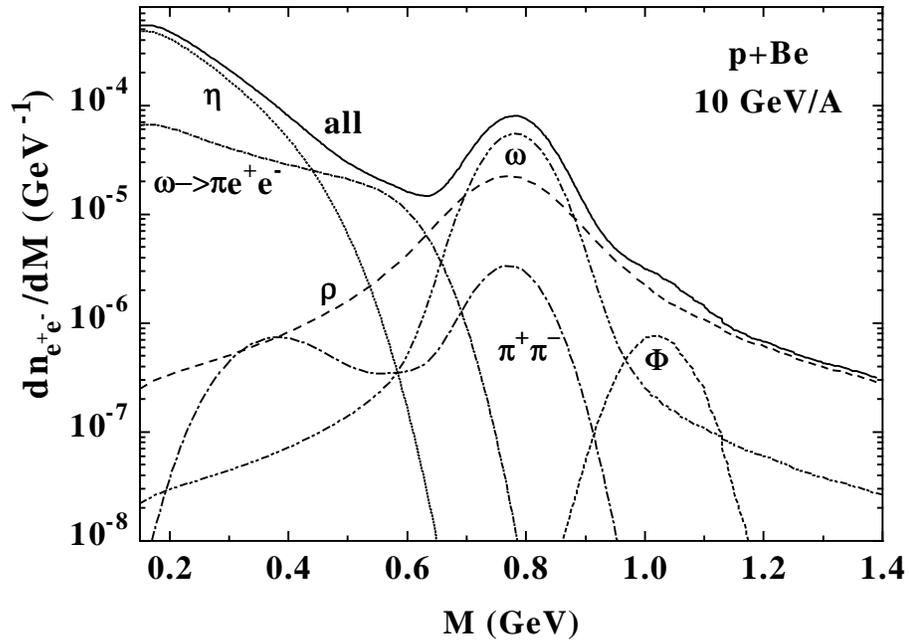
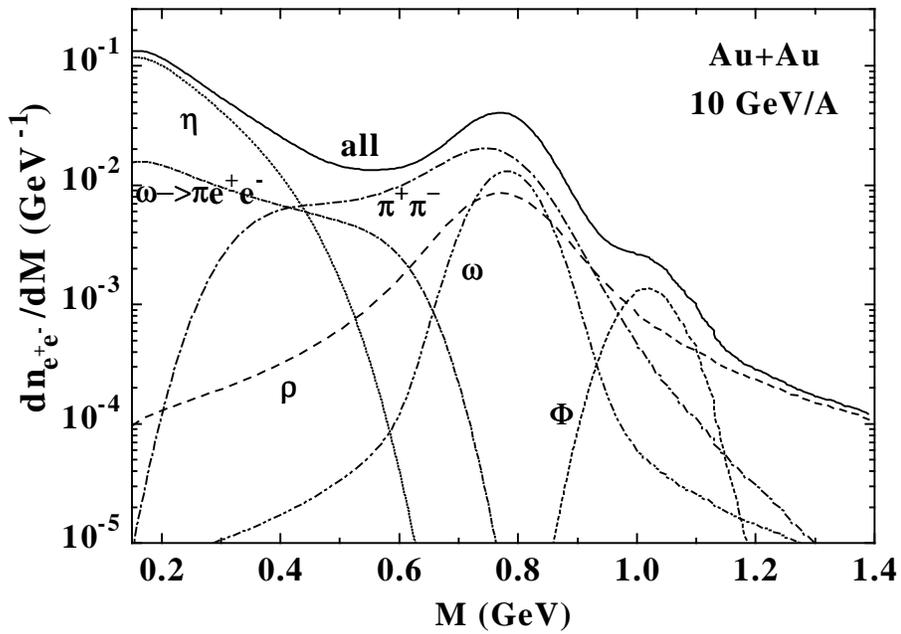
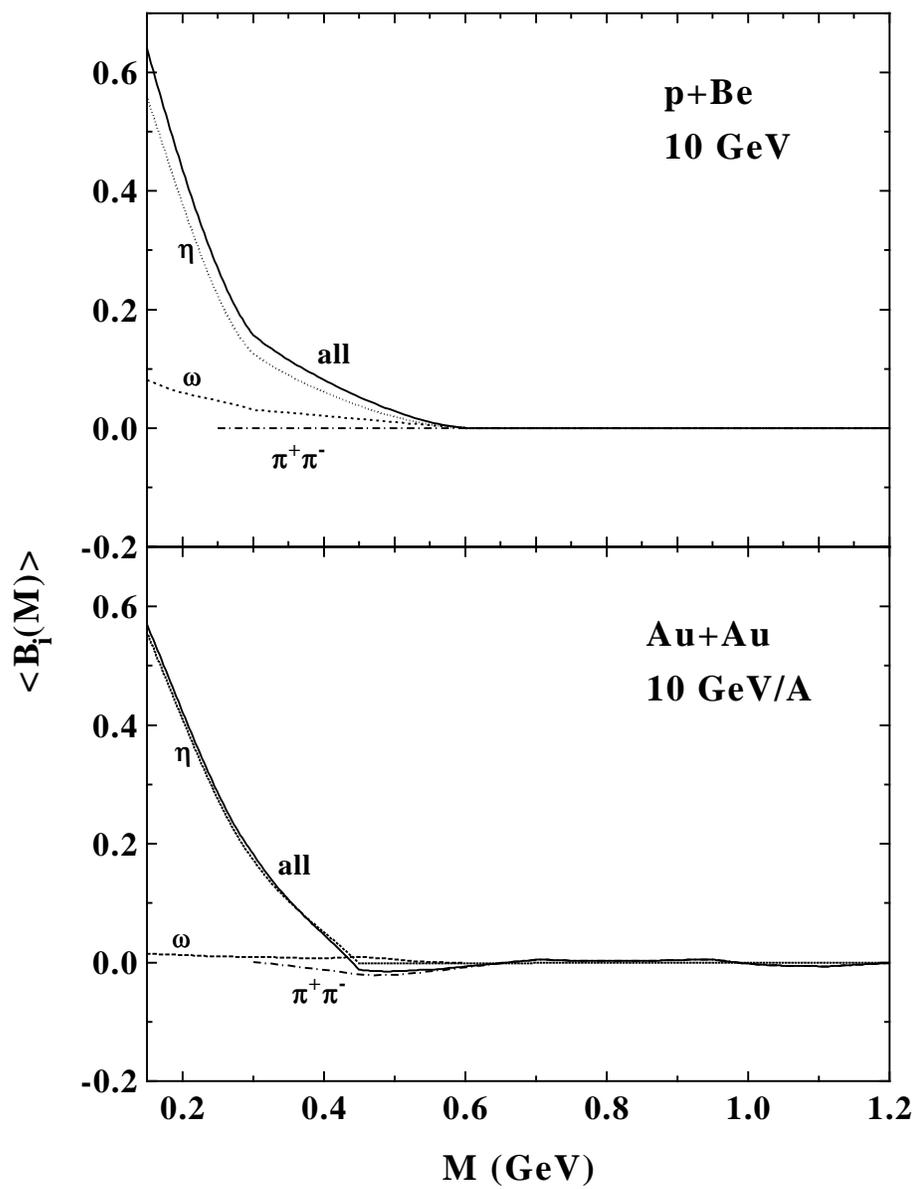
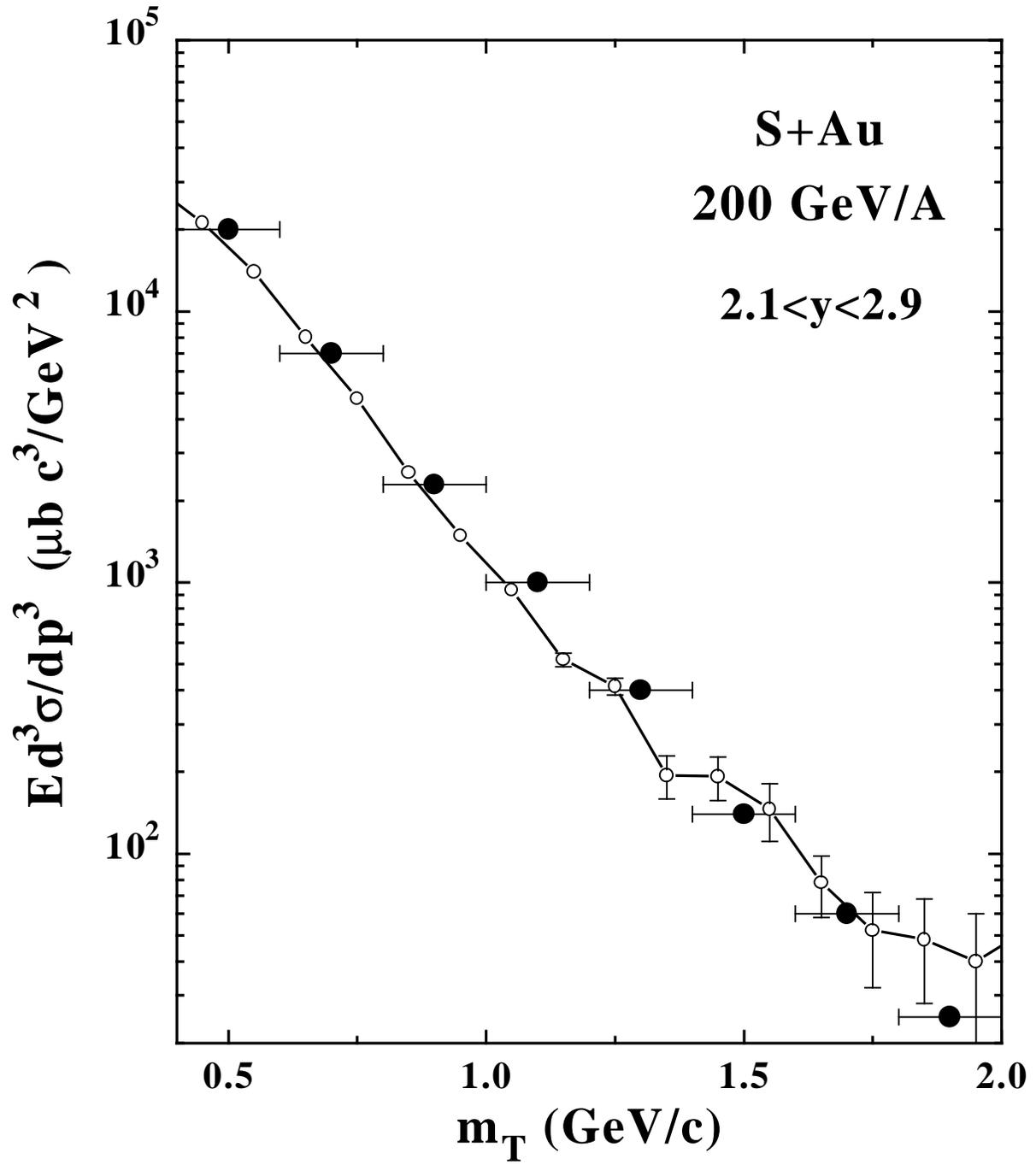
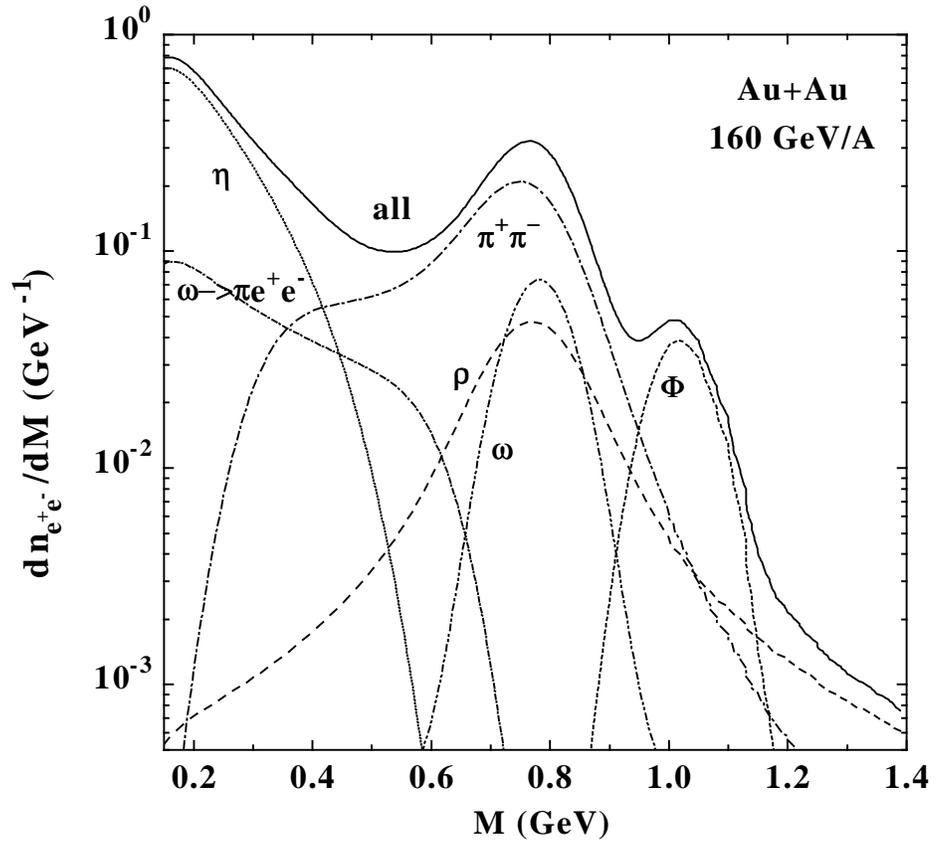
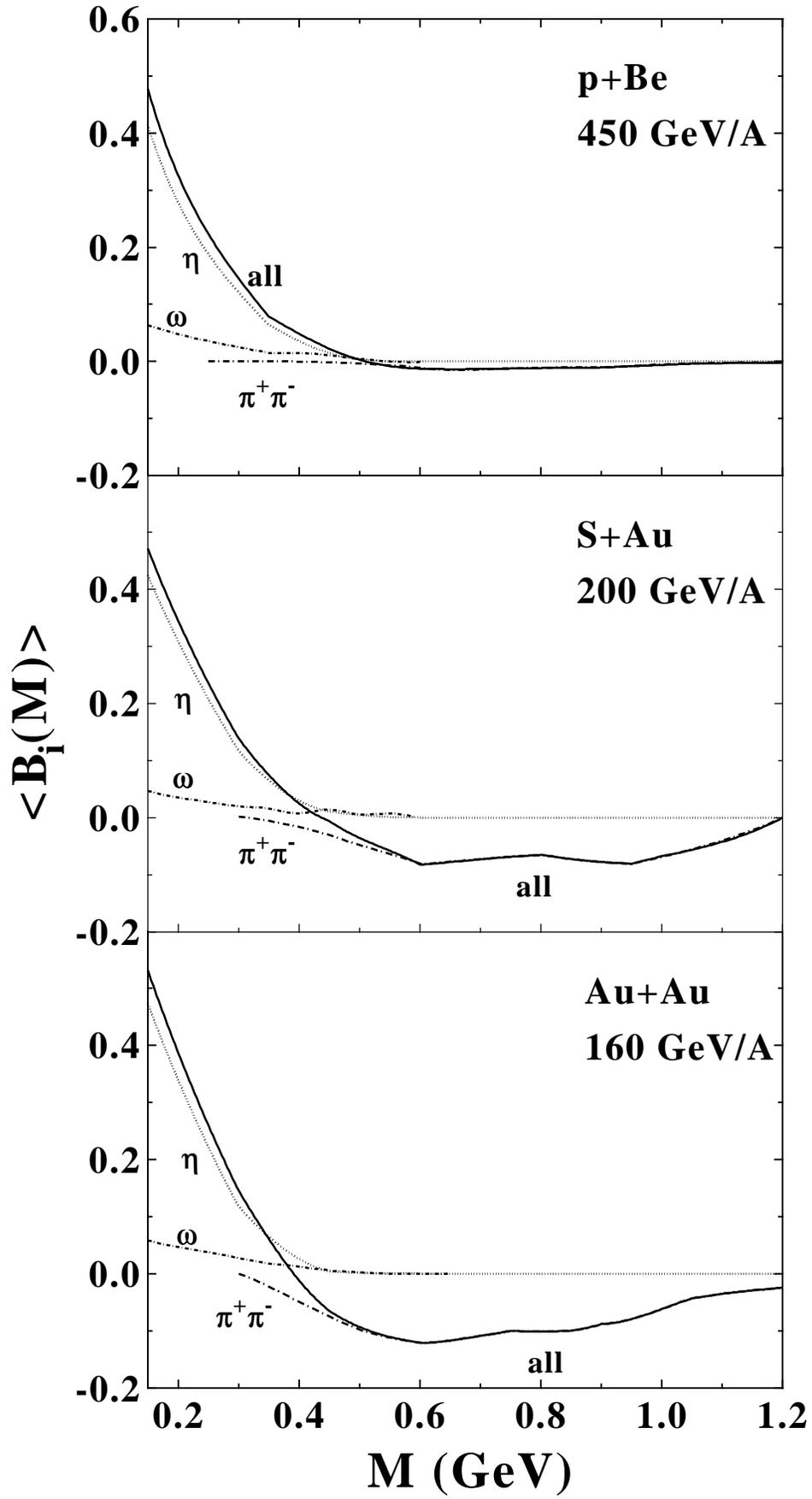
\end{document}